# Orbitally controlled quantum Hall states in decoupled two-bilayer graphene sheets

*Soyun Kim[1†], Dohun Kim[1†], Kenji Watanabe[2], Takashi Taniguchi[3], Jurgen. H. Smet[4], and Youngwook Kim[1*]*

[1]*Department of Physics and Chemistry, Daegu Gyeongbuk Institute of Science and Technology (DGIST), Daegu 42988, Republic of Korea*

[2]*Research Center for Functional Materials, National Institute for Materials Science, Tsukuba 305-0044, Japan*

[3]*International Center for Materials Nanoarchitectonics, National Institute for Materials Science, Tsukuba 305-0044, Japan*

[4]*Max Planck Institute for Solid State Research, Stuttgart 70569, Germany*

[†]These authors contributed equally

[*]E-mail: y.kim@dgist.ac.kr

**Abstract**

**We report on integer and fractional quantum Hall states in a stack of two twisted Bernal bilayer graphene sheets. By exploiting the momentum mismatch in reciprocal space, we suppress single particle tunneling between both bilayers. Since the bilayers are spatially separated by only 0.34 nm, the stack benefits from strong interlayer Coulomb interactions. These interactions can cause the formation of a Bose-Einstein condensate. Indeed, such a condensate is observed for half filling in each bilayer sheet. However, only when the partially filled levels have orbital index 1. It is absent for partially filled levels with orbital index 0. This discrepancy is tentatively attributed to the role of skyrmion/anti-skyrmion pair excitations and the dependence of the energy of these excitations on the orbital index. The application of asymmetric top and bottom gate voltages enables to influence the orbital nature of the electronic states of the graphene bilayers at the chemical potential and to navigate in an orbital mixed space. The latter hosts an even denominator fractional quantum Hall state at total filling -3/2. Our observations suggest a unique edge reconstruction involving both electrons and chiral *p*-wave composite fermions.**

Recent advances in twistronics [1] have enabled the exploration of an extremely diverse set of quantum phases. They include unconventional superconducting ground states, correlated insulating phases as well as integer and fractional Chern insulating states [2-18]. These phases form due to strong correlations in a flat band environment that possesses non-trivial topological characteristics and offers a pool of nearly degenerate electronic states. Such flat bands can be engineered in the absence of a magnetic field by twisting two layers with the so-called magic angle [2,3]. However, there is also a more traditional alternative route to end up with topologically non-trivial flat bands in 2D systems in the presence of a perpendicular magnetic field that breaks time reversal symmetry. Landau quantization produces Landau levels with a macroscopic degeneracy equal to the number of flux quanta that pierce the sample. They are characterized by their orbital index $N$ that acts as a Chern number and determines the shape of the wave function of the states the Landau level hosts. Correlations in this highly degenerate environment again produce a wide variety of ground states including for instance fractional quantum Hall states, Wigner crystal as well as bubble and stripe charge density wave phases [19].

Combining Landau levels from two different layers by placing two layers on top each other with a user-controlled twist between them adds a geometric degree of freedom to explore correlation physics in partially uncharted areas of parameter space. Since Landau levels already come as flat bands, the choice of the twist angle is not restricted to the magic angle. The most interesting in such double layer systems is physics from interlayer Coulomb interactions [20,21] whose importance and strength are governed by the ratio of the layer separation $d$ and the magnetic length $l_B$: $d/l_B$. The magnetic length serves as a measure of the Coulomb interaction strength. Strong tunneling due to layer coupling should be avoided to ensure that these interlayer interactions dominate. Bilayer systems carved out of conventional semiconductors are forced to focus on $2 > d/l_B > 1$ [22-30]. Entering the very interesting regime of lower $d/l_B$ ratios is difficult, because it is an extraordinary challenge to achieve decoupled layers, i.e. to suppress tunneling when the interlayer distance is very small. In van der Waals structures composed of two layers, layer decoupling can be easily achieved by twisting the layers with angles well above the magic angle. The layer spacing is on the sub-nanometer scale making much lower $d/l_B$ ratios accessible, thereby boosting the interlayer interactions. Recently, first studies in this regime have been explored in a twisted graphene bilayer [31,32] as well as in twisted $WSe_2$ [33]. This has facilitated the survival of excitons and other emerging quasiparticles up to unprecedented temperatures of tens of Kelvin, which has not been possible in conventional semiconductor based double layer systems [31-33].

Here, we focus on interlayer interaction physics in a system composed of two Bernal stacked bilayer graphene sheets, that are twisted with respect to each other by a target angle of 10° to suppress single particle tunneling. They are separated by a van der Waals gap of only 0.34 nm. Top and bottom gates enable separate tuning of the density of the two Bernal bilayers which themselves are composed of a pair of AB stacked graphene sheets. These gates also allow for the application of a vertical electric displacement field between the two bilayers, so that Landau levels of different orbital nature can be partially occupied in each bilayer. We report evidence for a particularly strong Bose Einstein exciton condensate as well as a composite integer and fractional quantum Hall state involving presumed $p$-wave paired composite fermions.

Such stacks of two 10° twisted Bernal bilayer graphene sheets were fabricated using a dry pick-up process based on an elvacite stamp [34]. Two devices were investigated and will be referred to as device D1 and D2. The key results are comparable, but how well the interaction related ground states appear varies among devices mainly related to contact quality issues and specific disorder details. Since very precise angular alignment is not required, straight edges from the bilayer flakes making up a device were used for angular orientation rather than the common cut- or tear-and-stack techniques in magic angle devices [35]. The twisted layers were encapsulated with hBN at the top and the bottom with an approximate thickness of 40 nm. Additional graphite layers with a thickness between 5 and 10 nm served as front and back gates and were operated in the circuit configuration as shown in the insert to Fig. 1a. The entire structure was covered with another protective hBN layer. The hBN layers were stacked such as to avoid Moiré superlattice effects with the bilayer graphene sheets [36].

Fig. 1a shows the longitudinal resistance as a function of the top-gate voltage for device D1. With decreasing temperature, the resistance peak at the charge-neutrality point increases rapidly and reaches a value as high as 55 k$\Omega$ at 1.5 K. This signals the existence of a gap at charge neutrality in the electronic band structure. Color maps of the longitudinal resistance and its derivative with respect to the top gate voltage are plotted in the parameter plane spanned by the top and back gate voltage in panel b and c, respectively. The dashed black line in Fig. 1b marks charge neutrality. These maps reveal that the high resistance at charge neutrality initially persists also for non-zero electric displacement field (white region in panel b), but then disappears for larger fields. In previous transport studies on two stacked graphene bilayers, the appearance of a gap at charge neutrality in the absence of a displacement field was attributed to crystal fields. [37]. Although crystal fields are typically not important in graphene, they cannot be ignored in this twisted double-bilayer configuration. They are present in van der Waals heterostructures due to the potential difference between different atomic species. The upper graphene sheet making up the top AB stacked bilayer is in proximity with a hBN layer, whereas the lower graphene sheet is near the upper graphene sheet of the second AB stacked bilayer. As a result, a perpendicular crystal field appears. Its strength has been estimated to be about 0.13V/nm [38]. The second AB stacked bilayer experiences a crystal field of comparable magnitude, but pointing in opposite direction in view of the overall layer sequence. It is this intrinsic polarization in both bilayers that causes the opening of a small band gap [37,38].

Contrary to what occurs in Bernal stacked bilayer graphene [39], a displacement field cannot induce a gap or strengthen an existing gap in this double bilayer configuration. Instead, the displacement field closes the crystal field induced gap, because it causes a relative shift between the bands of both bilayers. For sufficiently large displacement, the bands overlap, the density of states no longer drops to zero at charge neutrality and the resistance drops back to small values (Fig. 1b). The band structure of the system at charge neutrality for zero as well as non-zero electric field displacement is schematically illustrated in Fig. 1d and 1e for the top and back gate voltage pairs marked with colored circular dots in Fig. 1c.

Regions in the gate voltage parameter space that correspond to only electron, only hole or co-existing hole and electron charge carriers can be identified straightforwardly in Fig. 1h. It plots the Hall resistance $R_{xy}$ as a function of $V_{tg}$ and $V_{bg}$ for a field $B = 0.1$ T at $T = 1.5$ K. Blue regions correspond to electron accumulation and red regions to hole accumulation in both bilayers. The boundaries of these regions are demarcated with dotted lines. Here, only one of the bilayers is populated. These boundaries are also visible in the data plotted in Fig. 1c. The green and yellow dots in this panel mark voltage pairs where only the bottom and top layer is populated with holes, respectively. The other bilayer does not contribute a net population of charge carriers. The corresponding band structures are depicted in Fig. 1f and 1g. In between these boundaries with zero charge population in one of the bilayers, there is a regime of mixed conduction where the two bilayers host charge carriers of opposite sign. A recent study on samples composed of two twisted graphene bilayers with a small twist angle (about 2°) has reported evidence for Fermi surface nesting in this regime with co-existing holes and electrons [40]. Such features are however absent in our samples with much larger twist angles. They were intentionally selected to reduce the layer coupling and efficiently suppress interlayer tunneling by enforcing a larger momentum transfer near the K-point for a charge carrier to switch bilayer. This is detrimental for Fermi surface nesting. In the remainder we will focus on the regime of hole accumulation only, as the magnetotransport properties are better developed for holes than for electrons in our devices. Ideally, the quality of the transport features for electron and hole charge carriers should be equivalent. However, in real samples this is seldom (if ever) the case, an issue that has been observed and discussed on numerous occasions in transport studies [27-34]

Fig. 2a displays the four terminal conductance $\sigma_{xx}$ as a function of the total filling factor $\nu_{tot}$ and the displacement field $D/\varepsilon_0$ at $T = 1.5$ K and $B = 9$ T. $\nu_{tot}$ is defined as the ratio of the total density in the system and the degeneracy of a single Landau level without counting any additional degrees of freedom. Note that there are additional two-fold degeneracies due to the spin and valley degree of freedom for each Landau level similar as in monolayer graphene. The discrete energy spectrum for a Bernal stacked graphene bilayer is described by $E_N = \pm \hbar w_C \sqrt{N(N-1)}$, where $\hbar$ is the Planck constant, $w_c$ the cyclotron frequency ($w_c = eB/m$) and $m$ the effective mass. Since $E_N$ equals zero for both $N = 0$ and $N=1$, the zero energy Landau level has an extra two-fold degeneracy that is absent for all other Landau levels. Sweeping the total filling factor and the displacement field produces a rich set of quantum Hall features as seen in Fig. 2a. In Fig. 2b, we plot the line trace corresponding to zero displacement field ($D/\varepsilon_0 = 0$), i.e. for balanced populations in both bilayers. In a system composed of two bilayers that do not mutually interact we anticipate quantum Hall behavior in the balanced case only at total even integer filling when the two constituents have simultaneously condensed in the same integer quantum Hall state so that $\nu_{tot} = 2\nu_t = 2\nu_b$. Here, $\nu_t$ and $\nu_b$ correspond to the filling factor of the top and bottom bilayer, respectively. In the absence of any interactions, each bilayers would condense in an integer quantum Hall state only when $\nu_t$ or $\nu_b$ equals 4M with M = …-3,-2,-1,1, 2, …. due to the spin, valley and orbital state degeneracies. However, at sufficiently high magnetic field or density, Coulomb interactions lift all these degeneracies within a single Bernal stacked graphene bilayer. This is schematically illustrated for the zero energy Landau level in Fig. 2c. The valley, spin and orbital state degeneracy are successively removed. The order in which this occurs has been addressed in previous magneto-transport studies on single bilayer graphene [41-44]. Due to the degeneracy lifting, each bilayer will exhibit quantum Hall physics at all integer values, $\nu_t = \nu_b = $ …-3, -2, -1, 0, 1, 2, 3 …. We then anticipate integer quantum Hall behavior for the twisted double bilayer for $\nu_{tot} = $…-6, -4, -2, 0, 2, 4, …. Pronounced conductance minima indeed appear at even total integer filling in Fig. 2a and 2b. However, in addition, conductance minima of varying strength are also visible at some odd integer total fillings, such as for instance $\nu_{tot} = $ -1 and -3, but not $\nu_{tot} = $ -5 and -7. The longitudinal conductance $\sigma_{xx}$ and Hall conductance $\sigma_{xy}$ for device D2 is shown in Fig. 3 and comparable features as in device D1 are observed. In Fig. 3b it can be seen that conductance minima at odd integer fillings are accompanied by the corresponding plateau in the Hall conductance confirming that the minimum stem from quantum Hall incompressible behavior. Odd integer quantum Hall states in the double bilayer system with equal charge carrier population in each bilayer can only arise as a result of interlayer Coulomb interactions and can be accounted for by the formation of an excitonic Bose Einstein condensate [20-33]. Remains to understand why such condensates do only appear at $\nu_{tot}$ = -1 and -3, but not $\nu_{tot}$ = -5 and -7. It turns out that the orbital character, i.e. orbital index, of the completely filled Landau levels at the chemical potential is different for both pairs. Previous magneto-transport studies on single bilayer graphene have indicated that for filling range -2 < $\nu$ < 0, the orbital index N of the partially filled level is 1, whereas for the range -4 < $\nu$ < -2 it is 0. This is depicted in a cartoon-like fashion in the top panel of Fig. 2d and follows from the level diagram in Fig. 2c based on Ref. [41-44]. This filing factor ranges translate into orbital index 1 for -4 < $\nu_{tot}$ < 0 and 0 for -8 < $\nu_{tot}$ < -4 for the twisted graphene bilayer system under study here (lower panel Fig. 2d). Therefore, it is natural to attribute the distinct behavior for both pairs of odd integer total fillings (-1, -3 versus -5, -7) to the orbital quantum number of the Landau level states at the chemical potential that are partially occupied at these fillings. The odd integer quantum Hall states appear for partially filled $N = 1$ levels and not $N = 0$ levels. Studies on both twisted graphene and WSe$_2$ bilayers have shown that integer quantum Hall states associated with completely filled $N = 1$ levels are far more robust than for $N = 0$ levels. This implies that the lowest energy excitation from the $N = 1$ levels has a substantially higher energy than the lowest energy excitation from the $N = 0$ levels. In double layer systems, the lowest energy excitations come in two different types: a conventional single particle-hole excitation or a skyrmion/anti-skyrmion pair excitation. The skyrmion and anti-skyrmion carry each an elementary charge, but of opposite sign, and both possess a spin texture [21]. The energy to create a skyrmion/anti-skyrmion pair increases with an increase of the orbital quantum number of the Landau level involved [32,33,45], whereas the energy to generate a particle-hole pair decreases with an increase in the orbital quantum number. [32,33]. As a result, the lowest energy excitation for N = 0 levels are skyrmion/anti-skyrmion pairs and the energy is small. This implies that the thermal activation energy or gap of the related quantum Hall state is mall [32, 33]. Consequently, the odd integer quantum Hall states involving the N = 0 levels are weak. Experimentally, this has been confirmed in twisted WSe$_2$ monolayers. It is natural to attribute the weakness or absence of excitonic quantum Hall ground states for N = 0 levels in our samples to the availability of low energy skyrmion/anti-skyrmion pair excitations for these levels.

Thermal activation energies were extracted from measurements acquired on device D2 in a variable temperature insert. In Fig. 3**a**, $\sigma_{xx}$ is plotted as a function of the total filling factor for various temperatures in the absence of a displacement field, $D/\varepsilon_0 = 0$, and at a magnetic field of 14 T. As for device D1, well developed odd integer quantum Hall states are visible at $\nu_{tot}$ = -1 and -3 corresponding to the partial filling of levels with orbital index $N = 1$. These states are marked with black arrows. Fig. 3b shows the Hall conductance and plateaus indeed appear at these fillings. Fig. 3**c** summarizes the extracted thermal activation energy $\Delta$ for these $\nu_{tot}$ = -1 and -3 states as a function of the magnetic field using a fit to the Arrhenius expression, exp (−$\Delta/2k_B T$). For $B$ = 14 T, this analysis yields a gap of 14 K and 8 K for $\nu_{tot}$ = -1 and -3, respectively. These values are two orders of magnitude larger than reported gap values for Bose Einstein condensates in GaAs double layer systems [20-23]. The large values are a result of the enhanced interlayer Coulomb interaction strength due to the sub-nanometer layer separation and the effective suppression of electronic tunneling by the twist angle induced momentum mismatch. These results are consistent with the odd-integer quantum Hall states reported in twisted bilayer graphene and twisted WSe$_2$ devices [31-33].

In order to address fractional quantum Hall effect physics (FQHE), device D2 was unloaded and remounted in a dilution refrigerator with a base temperature of 30 mK. Unfortunately, the sample quality somewhat suffered during the transfer to this different cryogenic system. In order to bring out clearly any relevant features, we plot $d\sigma_{xx}/dV_{tg}$ in Fig. 4a. Usual odd denominator fractional quantum Hall states such as $\nu_{tot}$ = -4/3, -5/3, -7/3, and -8/3 are visible and the corresponding features in $d\sigma_{xx}/dV_{tg}$ vary little with the displacement field. In addition the fractional quantum Hall state at $\nu_{tot}$ = -13/5 as well as the $\nu_{tot}$ = -3/2 state with even denominator are present, but they only exist in a limited range of the experimentally accessible displacement field. The $\nu_{tot}$ = -13/5 state for instance appears when $D/\varepsilon_0 < -15$, while the $\nu_{tot}$ = -3/2 state develops when $D/\varepsilon_0$ drops below -65. This goes back to two issues. As we create a charge imbalance and the displacement field is varied, the orbital index of the partially filled levels at the chemical potential in each bilayer can change. This is illustrated for the simpler case of a single Bernal stacked bilayer graphene sheet in Fig. 4b for which the mixing of levels with different orbital indices and flipping of the orbital index of the level at the chemical potential with varying filling has been studied in some detail in the literature [41-44]. The orbital index of the partially filled level at the chemical potential equals 1 for $-2 < \nu < 0$ when $D/\varepsilon_0 = 0$. However, the orbital index changes from 1 to 0 at $D/\varepsilon_0 = -15$ in the filling factor range

-2 < ν < -1, while it remains 1 until $D/\varepsilon_0$ = -100 for -1< ν < 0. In the double bilayer graphene stack, the charge imbalance and the associated displacement fields together with the potential drop in the perpendicular direction can in addition cause crossings of Landau levels of the bilayers such that again the orbital index of the partially filled levels are affected. The orbital quantum number of the partially filled levels in turn plays a crucial role for the robustness or appearance of conventional as well as exotic fractional quantum Hall states, since with a change of the orbital index or when levels with different orbital indices mix the Haldane pseudopotentials that describe the inter-electron interactions are modified [19,34,46]. For instance, the p-wave paired composite fermion quantum Hall state at even denominator fractional filling, first reported in conventional semiconductor systems at filling 5/2 [47], serves as a good example. It emerges for a partially filled N = 1 level, but not N = 0, because it requires a net attractive interaction that only develops for the N= 1 case due to overscreening of the Coulomb interaction when forming composite fermions, i.e. topologically bound states of electrons and quantized vortices.[48] For the Bernal stacked bilayer case Fig. 4b summarizes the outcome of previous studies about what orbital quantum number is dominant for states at the chemical potential and which fractional quantum Hall effect states are anticipated to develop in the filling and displacement field parameter plane. This graph serves as the basis to construct a similar diagram for the samples composed of two twisted graphene bilayers. It is shown in Fig. 4c [41-44]. The crossing of Landau levels appears as a diamond shape such as the white area in Fig. 4c between total filling -1 and -3 -75 < $D/\varepsilon_0$ < -50. In this white area charge carriers redistribute among the levels that cross. This diamond shape can be discerned in the experimental data of Fig. 4a. Multiple other examples are visible in the experimental data of Fig. 2a. The three differently colored areas in Fig. 4c each correspond to a different pair of orbital quantum numbers for the partially filled levels of the two bilayers. Accordingly, they host different fractional quantum Hall states. When the orbital indices both equal 0 for the two bilayers when $D/\varepsilon_0$ < -15 and -3 < $\nu_{tot}$ < -2, conventional Jain states [46] appear at $\nu_{tot}$ = -8/3, -13/5 and -7/3. They result from the condensation of both bilayers in either two fractional quantum Hall states such as for instance $\nu_{tot}$ = -7/3 = -2/3 + -5/3 or an integer and a fractional quantum Hall state such as $\nu_{tot}$ = -7/3 = -1 + -4/3. When tuning the imbalance between the population of the two bilayers by changing $D/\varepsilon_0$, we may anticipate transitions to different pairs of quantum Hall states. However, this occurs smoothly and no sharp or clear signatures appear for such transitions. In the filling factor range $-2 < \nu_{tot} < -1$ for $-60 < D/\varepsilon_0 < -10$, the partially filled levels are both characterized by orbital index $N = 1$ and $\nu_{tot} = -4/3$ and $-5/3$ states can be detected. Higher order odd denominator fractional states such as the $\nu_{tot} = -7/5$ and $-8/5$ states however are not observed. The absence of these states is consistent with the nature of the inter-electron interactions in levels with orbital index $N = 1$ and also with earlier fractional quantum Hall studies on bilayer graphene. As can be seen from the graph in Fig. 4c, it is possible to enter a regime where the partially filled levels of the two layers have mixed orbital indices at higher displacement fields: $N=0$ for one layer and $N = 1$ for the other layer. Under these circumstances, even denominator fractional quantum Hall physics can emerge and indeed in the experimental data of Fig. 4a, a pronounced $\nu_{tot} = -3/2$ quantum Hall state appears for displacement fields $D/\varepsilon_0$ between $-60$ and $-80$. It is a compound state made up of a hole integer quantum Hall state with filling $-1$ and a $p$-wave paired composite fermion state with filling $-1/2$.

In summary, we have investigated integer and fractional quantum Hall states in a stack of two bilayers twisted by a 10° angle. The twist causes a large mismatch in reciprocal space between the momenta of the electronic states belonging to the different bilayers. As a result, single particle vertical tunneling is strongly suppressed and the physics is governed by interlayer Coulomb interactions. For balanced populations, these interactions can give rise to excitonic Bose-Einstein condensation at half filling of the partially filled levels producing odd integer quantum Hall physics. This ground state is observed for the first time in this twisted double graphene bilayer system for partially filled levels with orbital index 1. The same physics does not appear for partially filled $N=0$ levels, presumably because the low energy of skyrmion-anti-skyrmion pair excitation reduces the energy gap of the incompressible ground state for such levels. By fine-tuning the displacement electric field, a population imbalance is created and it is possible to enter a regime where Landau levels with different orbital indices are populated. Under these conditions, even-denominator fractional quantum Hall state can appear. Here, a $\nu_{tot} = -3/2$ state is observed. It is a compound state composed of the integer $\nu = -1$ state in one layer and the even-denominator fractional quantum Hall states $\nu = -1/2$ originating from p-wave pairing of composite fermions for the other layer.

**Acknowledgement**

This work was supported by the Basic Science Research Program NRF-2020R1C1C1006914, NRF-2022M3H3A1098408 through the National Research Foundation of Korea (NRF), the DGIST R&D program (22-CoE-NT-01) of the Korean Ministry of Science and ICT, and the BrainLink program funded by the Ministry of Science and ICT through the National Research Foundation of Korea (2022H1D3A3A01077468). We also acknowledge the partner group program of the Max Planck Society. J.H.S is grateful for financial support from the EU core 3 Graphene Flagship Program and the SPP 2244 of the DFG. K.W. and T.T. acknowledge support from JSPS KAKENHI (Grant Numbers 19H05790, 20H00354 and 21H05233) and A3 Foresight by JSPS.

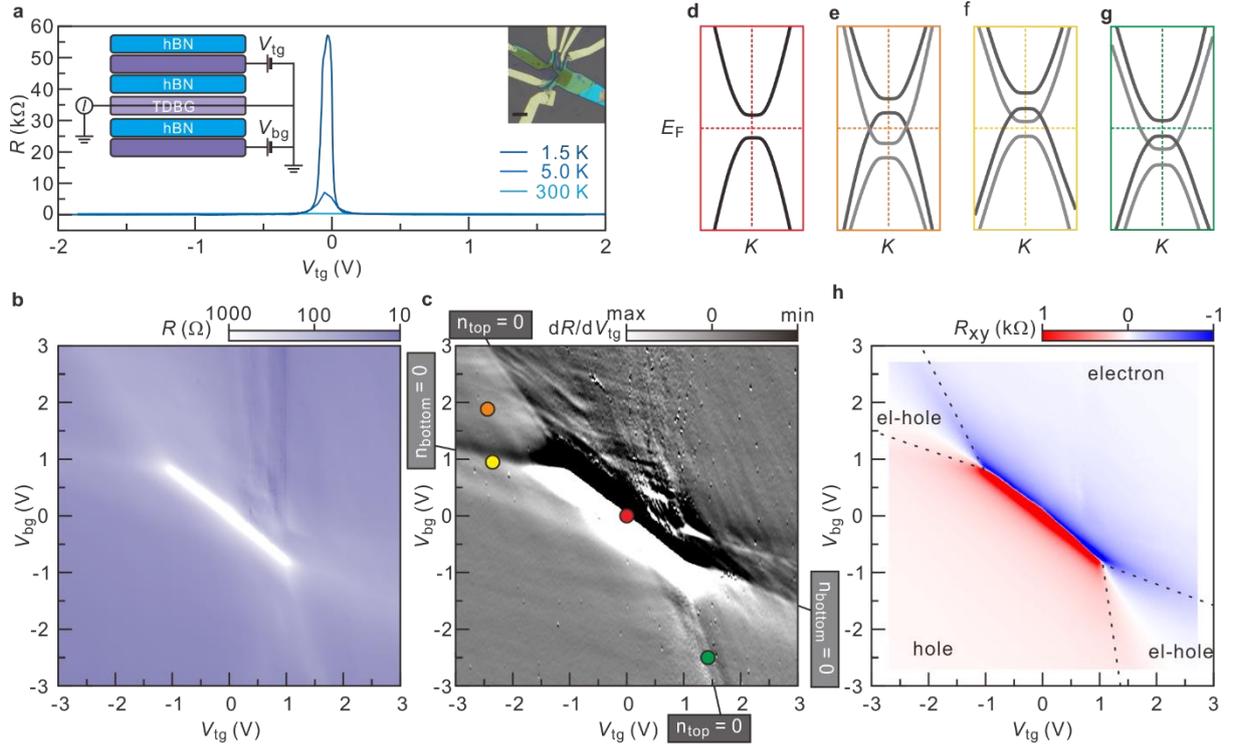

**Figure 1 a**, Longitudinal resistance as a function of the top gate voltage recorded at a temperature $T$ of 1.5, 5.0 and 300 K. The schematic diagram of the measurement circuit is displayed on the left inset. The right inset shows an optical image of device D1 (scale bar: 5 µm). **b**, Color map of the longitudinal resistance in the back gate and top gate voltage parameter plane for $T = 1.5$ K and $B = 0$ T. The dotted line marks overall charge neutrality: $n_{top} + n_{bottom} = 0$. **c**, Color map of $dR/dV_{tg}$ as a function of the top and bottom gate voltage. The dark and light grey boxes mark line features in the data set associated with charge neutrality (i.e. zero average density) in the top and bottom layer, respectively. **d-g**, Schematic of the band structure near the $K$-points in the Brillouin zone for different $(V_{tg}, V_{bg})$. The different color of the box surrounding each of these panels corresponds to the colored symbols in panel c that mark the relevant $(V_{tg}, V_{bg})$ pair. **h**, Color map of $R_{xy}$ as a function of top and bottom gate voltage. Blue and red colors indicate a net electron or hole population, respectively. The label "el-hole" denotes the mixed regime where electrons and holes co-exist.

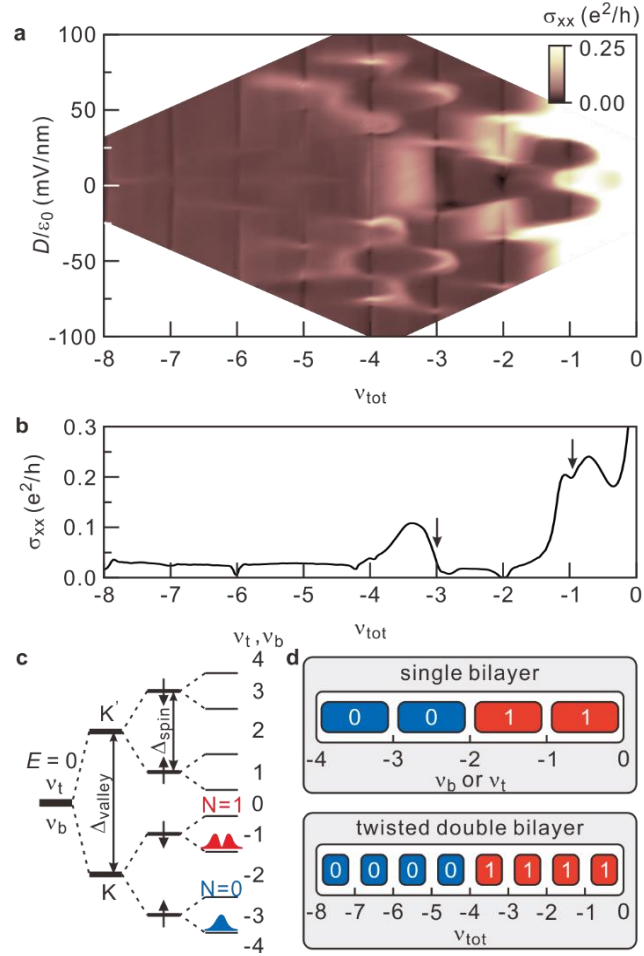

**Figure 2 a**, Color map for $\sigma_{xx}$ in the $(D/\varepsilon_0, \nu_{tot})$-plane. The data are recorded for $T = 1.5$ K and $B = 9$ T. **b**, Single line trace of $\sigma_{xx}$ as a function of $\nu_{tot}$ for $B = 9$ T and $D/\varepsilon_0 = 0$. Arrows mark the conductivity minima corresponding to odd total integer filling. **c**, Schematic of the splitting of the lowest Landau level at $D/\varepsilon_0 = 0$. This scheme does not take into consideration the Coulomb interaction between layers. **d**, Schematic of the evolution with total filling factor of the orbital index of the partially filled level at the chemical potential when $D/\varepsilon_0 = 0$. The top panel is for Bernal stacked bilayer graphene and the bottom panel for the stack of two twisted bilayer graphene sheets studied in this work.

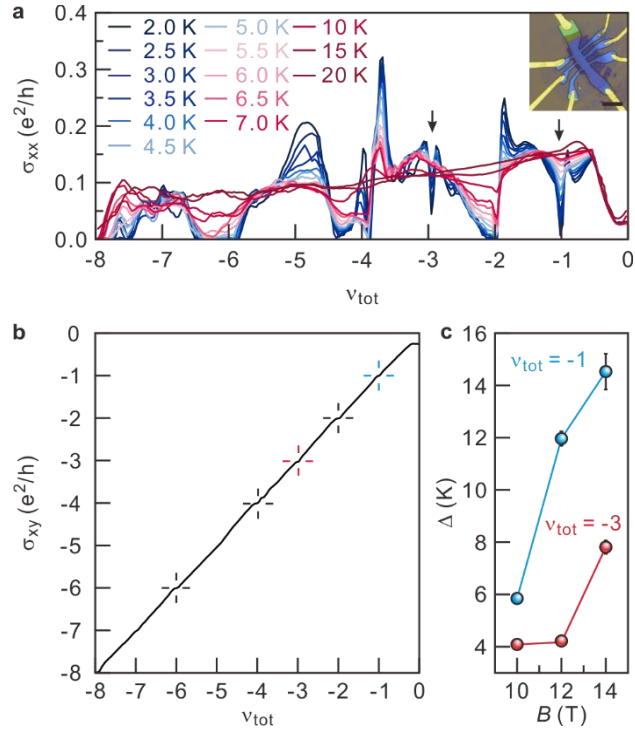

**Figure 3 a**, $\sigma_{xx}$ as a function of total filing factor at $T = 2$ K and $B = 14$ T. Each line trace is recorded at a different temperature and colored according to the legend. The two arrows mark odd-integer quantum Hall states at half filling of the partially filled level in each bilayer with orbital index $N = 1$. The insert is an optical image of device D2. **b**, $\sigma_{xy}$ as a function of total filing factor for $T = 2$ K and $B = 14$ T. Hall conductance plateaus at even integer total filling are highlighted with a black cross. Blue and red crosses mark the plateau at $\nu_{tot} = -1$ and -3, respectively. **c**, Thermal activation energy extracted from an Arrhenius fit of the temperature dependent conductance minima at total filling $\nu_{tot} = -1$ and -3.

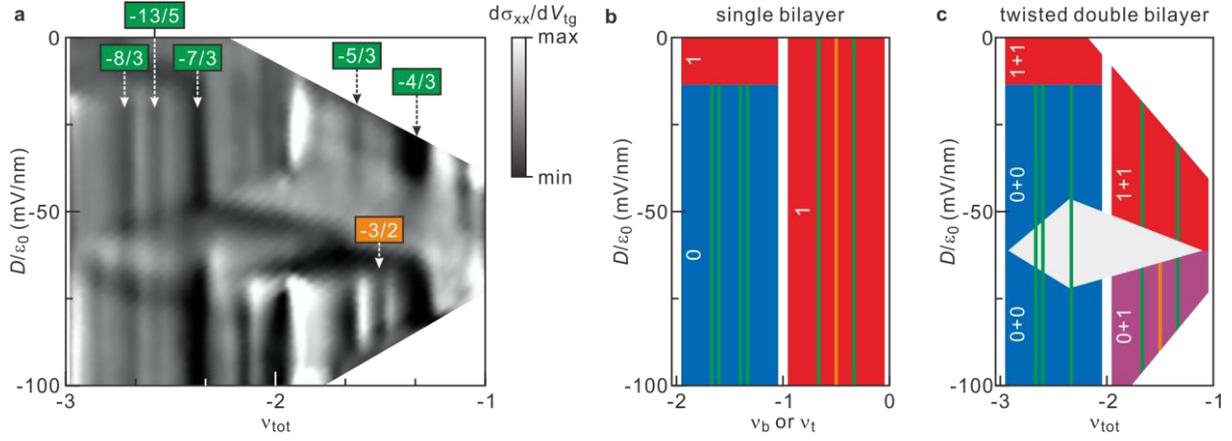

**Figure 4 a**, $dR_{xx}/dV_{tg}$ as a function of $D/\varepsilon_0$ and $\nu_{tot}$. Green boxes mark odd denominator fractional quantum Hall states, while the orange box highlights an even-denominator fractional quantum Hall state. **b**, Diagram showing the evolution of the orbital index of the partially filled level at the chemical potential in a Bernal stacked graphene bilayer as the filling factor $\nu$ and the displacement field are tuned. The red colored area is the parameter range where the orbital index equals 1, whereas blue is for N = 0. According to Ref [41-44], expected odd and even denominator fractional quantum Hall states are marked with green and orange lines. **c**, Same as in **b** but for a stack of two Bernal stacked bilayer graphene sheets. In the white area with diamond shape charge carriers are redistributed due to a crossing of Landau levels. In the purple area, two levels with different orbital character (N = 0 and N=1) are partially filled. The odd and even denominator fractional quantum Hall states observed in the data set plotted in panel **a** are shown as green and orange lines.